\documentclass{aa}

\usepackage{graphicx}
\usepackage{natbib}
\usepackage{siunitx}
\usepackage{nicefrac}
\usepackage{multirow}
\usepackage{pgfplots}
\usepackage[normalem]{ulem}
\usepackage[draft]{hyperref}
\usepackage{pgf,tikz}
\usepackage{verbatim}
\bibpunct{(}{)}{;}{a}{}{,}

\def \eos{EoS\xspace}
\def \amlt{\alpha_{\rm MLT}}
\def \afst{\alpha_{\rm FST}}
\def \cesampp{Cesam2k20\xspace}
\def \stagger{\textsc{stagger}\xspace}
\def \cobold{CO${}^5$BOLD\xspace}
\def \acen{$\alpha$\,Cen\xspace}
\def \feh{{\rm [Fe/H]}}
\def \sad{{s_{\rm ad}}}
\def \spresc{s_{\rm presc}}
\def \sbot{s_{\rm bot}}
\def \srhd{s_{\rm RHD}}
\def \presc{{\rm presc}}
\def \ds{{\rm d} s}
\def \teff{T_{\rm eff}}

\def \logg{\log g}
\def \erg{\mbox{erg}}
\def \K{\mbox{K}}
\def \sunit{10^9~\erg\,\mbox{g}^{-1}\,\mbox{K}^{-1}}
\def \punit{\mbox{dyn}\,\mbox{cm}^{-2}}
\def \lunit{~\erg\,\mbox{s}^{-1}\,\mbox{cm}^{-2}}
\def \rhounit{\mbox{g}\,\mbox{cm}^{-3}}
\def \gunit{\mbox{cm}\,\mbox{s}^{-2}}

\def \ttau{T(\tau)}

\newcommand{\corr}[1]{#1}

\newcolumntype{L}[1]{>{\raggedright\let\newline\\\arraybackslash\hspace{0pt}}m{#1}}

\usetikzlibrary{arrows,calc,fadings,decorations.pathreplacing,positioning,shapes,shadows,intersections,backgrounds,quotes,angles}
\tikzstyle{split_cell} = [rectangle, draw=black, anchor=north, rectangle split, rectangle split parts=2, align=center, text width=6cm]
\tikzstyle{myarrow}=[-latex', >=open triangle 90, thick]
\tikzstyle{line}=[-, thick]

\newcommand\st{\textsuperscript{st}\xspace}
\newcommand\nd{\textsuperscript{nd}\xspace}
\newcommand\rd{\textsuperscript{rd}\xspace}

\definecolor{louis}{HTML}{1f77b4}
\definecolor{federico}{HTML}{ff7f0e}
\definecolor{morgan}{HTML}{2ca02c}
\definecolor{mjo}{HTML}{d62728}
\definecolor{aldo}{HTML}{9467bd}
\definecolor{aldotwo}{HTML}{946700}
\definecolor{yveline}{HTML}{FF7E00}
\definecolor{hans}{HTML}{e377c2}
\definecolor{josefina}{HTML}{17becf}
\definecolor{laurent}{HTML}{bcbd22}

\begin{document}

    \title{Entropy-calibrated stellar modeling: Testing and improving the use of prescriptions for entropy of adiabatic convection}
    \titlerunning{Improvements of entropy-calibrated stellar modeling}

    \author{
        L.~Manchon\inst{1,2}
        \and M.~Deal\inst{3,4}
        \and M.-J.~Goupil\inst{2}
        \and A.~Serenelli\inst{5,6,7}
        \and Y.~Lebreton\inst{2,8}
        \and J.~Klevas\inst{9}
        \and A.~Ku\v{c}inskas\inst{9}
        \and H.-G.~Ludwig\inst{10}
        \and J.~Montalbán\inst{11,12}
        \and L.~Gizon\inst{1,13}}
    \institute{
    Max-Planck-Institut f\"ur Sonnensystemforschung, Justus-von-Liebig-Weg 3, 37077 G\"ottingen, Germany
    \and LESIA, Observatoire de Paris, PSL Research University, CNRS, Universit\'e Pierre et Marie Curie, Universit\'e Denis Diderot, 92195 Meudon, France\\
    \texttt{email: louis.manchon@obspm.fr}
    \and LUPM, Universit\'e de Montpellier, CNRS, place Eug\`ene Bataillon, 34095 Montpellier, France
    \and Centro de Astrof\'isica da Universidade do Porto, Rua das Estrelas, 4150-762 Porto, Portugal
    \and Institute of Space Sciences (ICE, CSIC), Carrer de Can Magrans S/N, 08193 Cerdanyola del Valles, Spain
    \and Institut d’Estudis Espacials de Catalunya (IEEC), Carrer Gran Capita 2, 08034 Barcelona, Spain
    \and Max-Planck-Institut f\"ur Astronomie, K\"onigstuhl~17, 69117 Heidelberg, Germany
    \and Universit\'e de Rennes, CNRS, IPR (Institut de Physique de Rennes) -- UMR 6251, 35000 Rennes, France
    \and Institute of Theoretical Physics and Astronomy, Vilnius University, Saul\.{e}tekio al. 3,  10257 Vilnius, Lithuania
    \and Zentrum f\"ur Astronomie der Universit\"at Heidelberg, Landessternwarte, K\"onigstuhl 12, 69117 Heidelberg, Germany
    \and Dipartimento di Fisica e Astronomia Augusto Righi, Universit\`a degli Studi di Bologna, Via Gobetti 93/2, 40129 Bologna, Italy
    \and School of Physics and Astronomy, University of Birmingham, Birmingham B15 2TT, UK
\and
Institut f\"ur Astrophysik, Georg-August-Universit\"at G\"ottingen, Friedrich-Hund-Platz~1, 37077 G\"ottingen}

    \date{Received XXX / Accepted XXX}

    \abstract{The modeling of convection is a long standing problem in stellar physics. Up-to-now, all \textit{ad hoc} models rely on a free parameter $\alpha$ (among others) which has no real physical justification and is therefore poorly constrained. However, a link exists between this free parameter and the entropy of the stellar adiabat. Prescriptions, derived from 3D stellar atmospheric models, are available that provide entropy as a function of stellar atmospheric parameters (effective temperature, surface gravity, chemical composition). This can provide constraints on $\alpha$ through the development of entropy-calibrated models. Several questions arise as these models are increasingly used.
    Which prescription should be used? Should they be used in there original form? How do uncertainties impact entropy-calibrated models?}
    {We aim to study in detail the three existing prescriptions and determine which one should be used, and how.}
    {We implemented the entropy-calibration method into the stellar evolution code (\cesampp) and performed comparisons with the Sun and the \acen system.{In addition, we used data from the CIFIST grid of 3D atmosphere models to evaluate the accuracy of the prescriptions.}}
    {Of three entropy prescriptions available, we determine which has the best functional form for reproducing the entropies of the 3D models.  However, the coefficients involved in this formulation must not be taken from the original paper, because they were calibrated against an improper set of entropies. We also demonstrate that the entropy obtained from this prescription should be corrected for the evolving chemical composition and for an entropy offset different between various EoS tables, following a precise procedure, otherwise classical parameters obtained from the models will be strongly biased. Finally, we also provide table with entropy of the adiabat of the CIFIST grid, as well as fits of these entropies.}
    {We performed a precise examination of entropy-calibrated modelling, and we gave recommendations on which adiabatic entropy prescription to use, how to correct it and how to implement the method into a stellar evolution code.}

    \maketitle

    \section{Introduction}
    The modeling of convection is a major problem for obtaining realistic stellar evolution models. Stellar convective regions are highly turbulent media where the Reynolds number can reach values of order $10^8$ (solar photosphere; \citealt{Komm1991}) to $10^{14}$ (bottom of the solar convective envelope; \citealt{Kupka2017}). In such regimes, Navier-Stokes equations can only be solved numerically. \corr{Conducting such hydrodynamic simulations would require high spatial resolutions on a large domain going from the top of adiabatic region to the outer layers of the photosphere, and such over stellar evolution time scale.} This is completely unreachable with the computational capabilities available presently and in the foreseeable future.

    In order to circumvent this problem, physicists have designed \textit{ad hoc} models that managed to reproduce the general properties of stellar convective envelopes, including their thermal structure. Among them, one can cite the Mixing Length Theory (MLT; \citealt{Bohm-Vitense1958}) or Full Spectrum of Turbulence models (FST; \citealt{Canuto1991,Canuto1992,Canuto1996}). The MLT reduces the spectral energy cascade of turbulence to a Dirac distribution, that is the whole convective flux is carried by eddies of a single size. A hot parcel of matter, before dissolving into the surrounding medium, rises a distance (called the mixing length) $\ell = \amlt H_p$, where $H_p$ is the pressure scale height and $\amlt$ is a free parameter. In deeper regions of the star, convective energy transport is very close to adiabatic, that is isentropic, the thermal structure is then independent of the convection formulation, and in particular of $\amlt$, used in the models \citep{Gough1976}. However, when the adiabaticity assumption breaks down, typically in low-density near-surface layers in convective stellar envelopes, the value of $\amlt$ determines the degree of non adiabaticity in the model, that is the entropy difference between the minimum of entropy occurring at the top of the convective envelope and the fully adiabatic interior. The FST represents an attempt to work around this dependence. It assumes a more complex form for the cascade, e.g. a Kolmogorov spectrum. Original formulations of the FST  models do not introduce free parameters and define the mixing length simply as the distance $z$ to the nearest boundary of the convective zone. For fine-tuning purposes, \citet{Canuto1996} introduced a dependence on the pressure scale height through a free parameter $\afst$, supposed to be small and to experience only small changes from a model to another.

    However \citet{Ludwig1999} found, using the thermal structures obtained from 2D radiation hydrodynamics (RHD) stellar atmosphere models, that $\afst$ does not remain small and varies by a similar amount as $\amlt$ over the HR diagram. Over the past decades, there have been many works highlighting the successes and caveats of MLT and FST models (e.g. \citealt{Arnett2015,Trampedach2010,Kupka2017}) and trying to improve or to go beyond them (e.g. \citealt{Gough1977,Grigahcene2005,Li2012,Pasetto2014,Gabriel2018}).

    Nonetheless, MLT and FST models remain widely used in stellar evolution codes. The most practical issue these models raise is: how do we choose the value of the free parameter $\alpha$\footnote{In the following, we use $\alpha$ equivalently for $\amlt$ or $\afst$, except when explicitly mentioned otherwise.}? There exists in general two ways to choose the $\alpha$ value. Firstly, through an optimization process, $\alpha$ can be tuned so that stellar models match a set of observable constraints for one or several stars. The most straightforward case is the widely used solar calibration, in which $\alpha$ is obtained by adjusting its value -- together with the chemical composition -- of models to reproduce the present-day global solar properties \citep[e.g.][]{Christensen-Dalsgaard1982,Morel2000,Serenelli2016,Vinyoles2017}. The same $\alpha$ value is then used to compute stellar models. This assumes that a unique value of $\alpha$ is appropriate for all stars at all evolutionary stages but there is a priori no reason why this should be the case \citep{Ireland2018} and RHD simulations of convective stellar atmospheres consistently point to a variable $\alpha$ \citep{Trampedach2014,Sonoi2019}. A possible extension would be to use stars, or sets of them, as calibrators, ideally establishing a semi-empirical relation between $\alpha$ values and stellar fundamental parameters. There is, however, a lack of reliable calibrators for which stellar parameters are known with high enough accuracy.

    Secondly, a different approach is to consider $\alpha$ as a completely free parameter that can be optimized on star-by-star basis to find best-fitted stellar models to a set of observables. This can be done using either stellar models that are computed as part of the optimization process or pre-computed grids of models with different $\alpha$-values together with an interpolation algorithm (e.g. SPInS; \citealt{Lebreton2020}) to find the best-fit model. This is actually not a calibration, as the resulting $\alpha$ values are not used as the basis for modeling other stars, only to obtain a best-fit model. However, such an approach gives little hope of improving on our formulations and understanding of convection.

    To improve the use of MLT or FST modeling of convection, one must find a relationship between $\alpha$ and global stellar parameters (effective temperature, surface gravity, chemical composition). This is the aim of $\alpha$ calibrations. \citet{Ludwig1999} used standard 1D stellar atmosphere models with a tunable $\amlt$ parameter (similar to what we have in stellar evolution models) and a grid of more realistic two-dimensional (2D) radiative-hydrodynamics stellar atmosphere models, to find a relation between $\amlt$, $\teff$ and $\logg$. More precisely, they adjusted the value of $\amlt$ so that the specific entropy of the adiabat of their envelope models matches the one of their 2D models. This work provided an expression for $\amlt$ as a function of effective temperature $\teff$ and surface gravity $g$ of the models. Aiming at improving this pioneering study, similar work was carried out by \citet{Trampedach2014} using solar-metallicity 3D atmosphere \corr{simulations computed with the \citet{Nordlund1990} code} and later by \citet{Magic2015} with 3D \stagger models with metallicity $\feh \in [-4.0; +0.5]$. More recently, \citet{Sonoi2019} used a grid of 3D stellar atmosphere models computed with \cobold (COnservative COde for the COmputation of COmpressible COnvection in a BOx of $L$ Dimensions with $L=2,3$; \citealt{Freytag2002,Freytag2012,Wedemeyer2004}) to obtain a representation of $\alpha$, as a function of global stellar parameters, for MLT and FST modeling, as well as for different temperatures $T$ vs. optical depth $\tau$ relations ($T(\tau)$ relations)  used to describe the atmosphere. All those prescriptions for $\alpha$ are very easy to implement in a stellar evolution code. However, they are highly dependent on the physical ingredients, as it is shown by the important change of calibrated $\alpha$ values introduced by a change of $T(\tau)$ relation.

    At the same time, \citet{Tanner2016} suggested to directly adjust the $\alpha$ parameter in stellar evolution calculations along the evolution such that the specific entropy of the model adiabat $\sad$ matches at all times $\srhd$, the value obtained in RHD simulations for the same $\teff$, $\log{g}$ and $\feh$ values (notations for the various specific entropies are summarised in Table \ref{table:s_notation}). This has the advantage that the specific entropy has a well defined physical meaning and is linked to the surface properties of the star by means of the realistic RHD models. In this way, $\alpha$ simply becomes a parameter used to ensure that stellar models reproduce the correct $\srhd$, as inferred from RHD simulations, in a consistent way with the physical ingredients in the stellar evolution models, e.g. different $\ttau$ relations, use of MLT, FST or any other prescription for convection. The idea of \citet{Tanner2016} was implemented in YREC (Yale Rotational stellar Evolution Code; \citealt{Demarque2008}) and tested on the Sun \citep{Spada2018}, on the \acen system \citep{Spada2019} and the method was finally extended to evolved stars \citep{Spada2021}.

    The aim of the present work is to contribute to the general effort to develop more accurate stellar models. Modelling the atmosphere of solar-type stars is based on three main components : \textit{(1)} the opacity profile in the atmosphere, \textit{(2)} the $\ttau$ relation which defines the transition in the temperature gradient between optically thick and optically thin regions, and \textit{(3)} the convection formalism. For a seismically reliable stellar model, these three components must be chosen with the utmost care. This work focuses on the last component by removing the main free parameters of the convection formalism. Specifically here, we built on previous works to produce stellar models which no longer include the mixing length as a free parameter. To that purpose  we implemented the same method in the stellar evolution code \cesampp (formerly known as CESTAM: Code d'Evolution Stellaire avec Transport, Adaptatif et Modulaire; \citealt{Morel1997,Morel2008,Marques2013}), as in the YREC code \citep{Demarque2008}. The previous works made it clear that a lot of care must be taken in the use of entropy prescription. In Sect. \ref{section:sec2}, we review the available prescriptions and in Sect. \ref{section:sec3}, we describe the implementation of entropy calibration in \cesampp. In Sect. \ref{section:sec4} we describe the necessary improvements that should be added to the prescriptions. A discussion is given in Sect. \ref{section:sec5} and conclusion in Sect. \ref{section:sec6}.

    \section{Available prescriptions for the entropy at the bottom of the convective envelope}
    \label{section:sec2}

    \begin{table}
        \caption{Summary of the different notations used to denote the specific entropy.}
        \label{table:s_notation}
        \centering
        \begin{tabular}{l L{0.8\linewidth} }
            \hline\hline
                $\sad$    & Specific entropy of the adiabat in 1D stellar evolution models.\\[7pt]
                $\srhd$   & Specific entropy of the adiabat in 3D stellar atmosphere models. Entropy assigned to the inflow at the bottom.\\[7pt]
                $\spresc$ & Value of the specific entropy of the adiabat, obtained from a prescription. \\[7pt]
                $\sbot$   & Average specific entropy at the bottom layer of 3D simulations of stellar atmospheres. \\[7pt]
            \hline
        \end{tabular}
    \end{table}

    \subsection{Stellar atmosphere codes}

    Before describing the entropy prescriptions, it is worth giving some details about the 3D radiation-hydrodynamics model atmosphere codes that were used to compute the simulations on which they are based. The atmosphere codes can solve hydrodynamic and radiative transfer equations. All simulations used hereafter do not include the magnetic field and therefore are referred to as RHD simulations. The codes rely on the "box-in-a-star" setup, where only the flow included in a box is simulated. This box covers an optical depth range of $-6 \leq \log_{10}\tau_{\rm Ross} \leq 5$, which encompasses the superadiabatic region. This domain is small in comparison to stellar dimensions but large enough to resolve properties of the convective flow. Periodic boundary conditions are assumed at the vertical faces of the box and, at horizontal faces, the flow is free to enter and leave the box. The upper boundary condition can vary from one code to the other. At the bottom, the specific entropy of the in-flowing material is fixed which allows to constrain the effective temperature of the resulting model. The specific entropy of the inflow can be seen as the entropy of the adiabat. Because at the lower boundary, inflow and outflow do not have the same entropy, the horizontally average at this depth is not equal to the entropy of the adiabat.

    In the following, we always denote $\spresc$ the specific entropy obtained from a prescription, $\srhd$ the specific entropy of the adiabat of an atmosphere model, and we use the word 'entropy' as a synonym for 'specific entropy' (entropy per unit of mass).
    Up-to-now, three different mathematical formulations have been proposed to relate $\srhd$ to $\teff$, $\log g$ and, optionally, $\feh$. While their general mathematical forms remain similar, they have been calibrated on different sets of models.

\begin{figure}
\begin{center}
\begin{tikzpicture}
    \node (input)      [           align=center, node distance=1mm                                                                            ]{\textit{0. At time $t+\Delta t$}};

    \node (limzc)      [rectangle, align=center, draw=black, below=of input,      text width=4.2cm, yshift=0.5cm                              ]{1. Find limit RZ/CZ};
    \node (converge_1) [rectangle, align=center, draw=black, below=of limzc,      text width=2.8cm, yshift=0.5cm, minimum size=8mm, anchor=north]{\small \textit{Convergence ?}};

    \node (chim)       [rectangle, align=center, draw=black, below=of converge_1, text width=4.2cm, yshift=0.5cm                              ]{2. Evolution of chemical composition};
    \node (converge_2) [rectangle, align=center, draw=black, below=of chim,       text width=2.8cm, yshift=0.5cm, minimum size=8mm, anchor=north]{\small \textit{Convergence ?}};

    \node (QS_1D)      [rectangle, align=center, draw=black, below=of converge_2, text width=4.2cm, yshift=0.5cm                              ]{3. Compute new quasi-static structure};
    \node (converge_3) [rectangle, align=center, draw=black, below=of QS_1D,      text width=2.8cm, yshift=0.5cm, minimum size=8mm, anchor=north]{\small \textit{Convergence ?}};

    \node (output)     [           align=center,             below=of converge_3,                 yshift=0.5cm]                                {\textit{4. Next time step}};

    \draw[myarrow] (input.south)      --                    (limzc.north);
    \draw[myarrow] (limzc.south)      --                    (converge_1.north);
    \draw[myarrow] (converge_1.east)  -- node[above]{no} ++ (1.5,0) |- (input.east);
    \draw[myarrow] (converge_1.south) -- node[right]{yes}    (chim.north);

    \draw[myarrow] (chim.south)       --                    (converge_2.north);
    \draw[myarrow] (converge_2.east)  -- node[above]{no} ++ (1.5,0) |- node[right]{\centering $\Delta t = \frac{\Delta t}{2}$} (input.east);
    \draw[myarrow] (converge_2.south) -- node[right]{yes}    (QS_1D.north);

    \draw[myarrow] (QS_1D.south)      --                    (converge_3.north);
    \draw[myarrow] (converge_3.east)  -- node[above]{no} ++ (1.5,0) |- (input.east);

    \draw[myarrow] (converge_3.west)    -- node[above]{no} ++ (-2,0) |- node[above]{\centering re-iterate}  (limzc.west);
    \draw[myarrow] (converge_3.south)   -- node[right]{yes} (output.north);
\end{tikzpicture}
\caption{Schematic representation of the step of a Newton-Raphson scheme followed by \cesampp during the computation over a time step. "Problem" can be an unconverged solution, an interpolation outside EoS or opacity table, }
\label{fig:cest_time_step}
\end{center}
\end{figure}
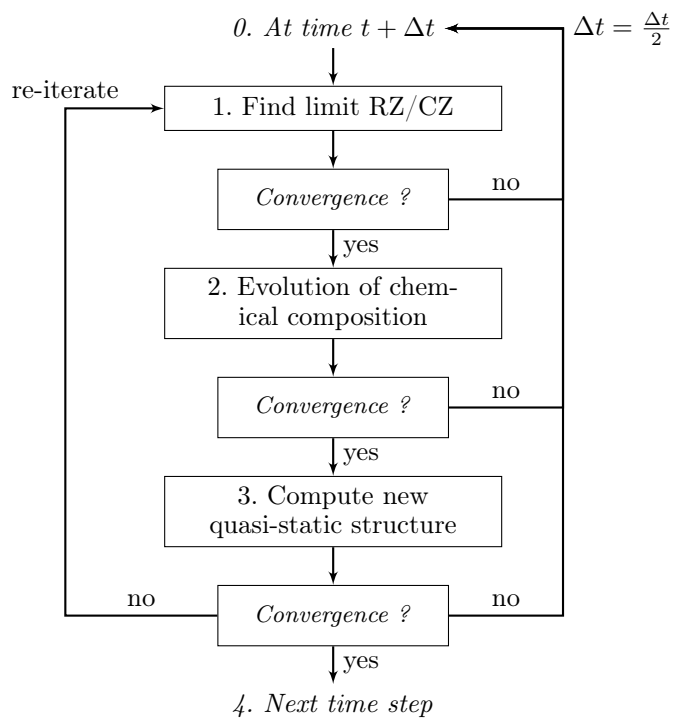

    \subsection{\citet{Ludwig1999}'s functional form}

    The original idea that models of surface convection could be used to derive constraints on the value of the entropy of the adiabat was first suggested by \citet{Steffen1993}. \citet{Ludwig1999} (L99 in the following) used a grid of 58 two-dimensional RHD atmosphere models with $4300~\mbox{K} \leq \teff \leq 7100~\mbox{K}$, $2.54 \leq \log g \leq 4.74$ and a chemical composition close to that of the proto-Sun composition proposed by \citet{Grevesse1998} (hereafter referred to as GS98), with $Y=0.28$ and $Z=0.016$. The entropy of the adiabat is expressed as follows:
    \begin{equation}
        \spresc^{\rm L99}(\widetilde T, \widetilde g) = a_0 + a_1 \widetilde{T} + a_2 \widetilde{g} + a_3\exp( a_4 \widetilde{T} + a_5 \widetilde{g} ),
        \label{eq:s_lud}
    \end{equation}
    where $\{a_i\}_{i=0}^5$ are constants determined using  the grid of models, $\widetilde{T} = (\teff - 5770)/1000$ and $\widetilde{g} = \log( g / 27500)$.

    \subsection{\citet{Magic2013}'s functional form}

    \citet{Magic2013} (M13 in the following) used a grid (the \stagger-grid) of 217 3D RHD atmosphere models computed with the \stagger code \citep{Nordlund1995}, with $4000~\mbox{K} \leq \teff \leq 7000~\mbox{K}$ in steps of $500~\mbox{K}$, $1.5 \leq \log g \leq 5.0$ in steps of $0.5$ and 7 values of metallicity $\feh = +0.5, 0.0, -0.5, -1.0, -2.0, -3.0, -4.0$. The solar chemical composition follows the determination of \citet{Asplund2009} (hereafter denoted AGSS09), for the present Sun. M13 suggested to represent the entropy of the adiabat with the  following function:
    \begin{align}
        \spresc^{\rm M13}(\widetilde T, \widetilde g, \widetilde z) = &P_2(a_i, \widetilde z) + P_2(b_i, \widetilde z) \widetilde{T} + P_2(c_i, \widetilde z) \widetilde{g} \nonumber\\
                                                               &+ P_2(d_i, \widetilde z)\exp( P_2(e_i, \widetilde z) \widetilde{T} + P_2(f_i, \widetilde z) \widetilde{g} ),
        \label{eq:s_mag}
    \end{align}
    and
    \begin{equation}
        P_2( a_i, \widetilde z) = \sum_{i=0}^2 a_i \widetilde z^i,
        \label{eq:P2}
    \end{equation}
    where the coefficients $\{a_i, b_i, c_i, d_i, e_i, f_i\}_{i=0}^2$ are constants, tuned to fit the entropy $\srhd$ of the \stagger-grid models, $\widetilde T = (\teff - 5777)/1000$, $\widetilde g = \log g - 4.44$, $\widetilde z = \feh$. We stress that $\widetilde z$ must be computed with respect to an AGSS09 chemical mixture. If one wants to use another chemical composition of reference, the coefficients in Eq. \eqref{eq:s_mag} should be recomputed. The expression proposed by M13 is essentially the same as the one of L99, except that the M13 one takes into account the metallicity.

    \subsection{\citet{Tanner2016}'s functional form}

    Based on the same \stagger-grid, \citet{Tanner2016} proposed a different form to represent $\sad$ as a function of $\teff$, $\logg$ and $\feh$:
    \begin{equation}
        \spresc^{\rm T16}(\widetilde T, \widetilde g; \widetilde z) = s_0 + \beta \exp\left( \frac{A\widetilde T + B \widetilde g - x_0}{\tau_0} \right),
        \label{eq:s_tan}
    \end{equation}
    where $s_0$, $x_0$, $\beta$, $\tau_0$,  $A$ and $B$ are constants, calibrated for each sub-set of models with same metallicity; $\widetilde T = \log \teff$ and $\widetilde g = \log g$.

    \section{Implementation of entropy-calibration in \cesampp}
    \label{section:sec3}

    \begin{table}[t]
        \caption{Optimal parameters obtained for the standard solar model SSM calibrated with the OSM (Optimal Stellar Model) stellar optimization tool.}
        \label{table:optimal_sun}
        \centering
        \begin{tabular}{c c c }
            \hline\hline                                                                            \\[-6pt]
            \multicolumn{3}{c}{Observational constraints:} \\
            $L~[L_\odot]$      & $\teff~[\K]$    & $(Z/X)_{\rm s}$                                  \\[3pt]
            $1$  & $5772$  & $0.0231$                                                               \\[5pt]
            \multicolumn{3}{c}{Adjustable parameters:}       \\
            $Y_0$                          & $(Z/X)_0$              & $\amlt$                       \\[3pt]
            $0.27334 \pm 4\cdot 10^{-5} $  & $0.02651 \pm 10^{-5}$  & $1.828 \pm 3\cdot 10^{-3}$    \\
            \hline
            \end{tabular}
    \end{table}

    \subsection{Standard \cesampp}

    \begin{figure}[t]
        \begin{center}
            \includegraphics[draft=False]{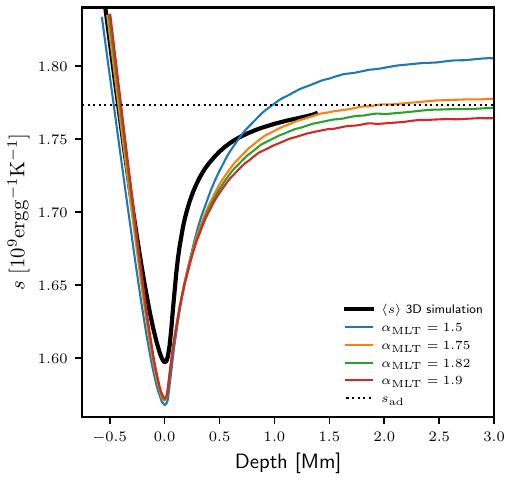}
            \caption{Profile of specific entropy as a function of the depth in the model. The origin of the depth is taken as the location of the \corr{entropy} minimum in the superadiabatic region. Average specific entropy profile of a 3D simulation is represented as a solid black line, and the entropy of the adiabat as a dash black line. The entropy of the adiabat is an input of the 3D model. Other curves are profiles of specific entropy obtained with 1D models computed with different values of $\amlt = 1.5$ (blue), $1.75$ (orange), $1.82$ (green) and $1.9$ (red).}
            \label{fig:alpha_s_match}
        \end{center}
    \end{figure}

    The \cesampp code\footnote{\url{https://www.ias.u-psud.fr/cesam2k20/}} implements the same numerical methods as its predecessors (for details, see \citealt{Morel1997} for CESAM, \citealt{Morel2008} for CESAM2k and \citealt{Marques2013} for CESTAM). Stellar structure equations are solved using a collocation method where solutions are represented as a linear combination of piecewise polynomials, projected over a B-Spline basis.

    \cesampp uses opacity tables either from the OPAL team \citep{Rogers1992,Iglesias1996} or OP tables from the Opacity Project \citep{Seaton1994,Badnell2005,Seaton2005,Seaton2007}. Several equations of state (\eos) are implemented and this work uses tables from the OPAL5 \eos \citep{Rogers2002}. Concerning the nuclear reaction rates, we used compilations from NACRE \citep{Aikawa2006} (except LUNA \citep{Broggini2018} for the ${}^{14}{\rm N(p},\gamma){}^{15}{\rm O}$ reaction).

    Regarding the mixing processes, rotation is neglected throughout this work. The effects of atomic diffusion are modelled with the formalism proposed by \citet{Michaud1993}. Convection is modelled with mixing-length theory formalism under the formulation of \citet{Henyey1965} taking into account the optical thickness of the convective bubble. We stress that the entropy-calibration method also works with more complex formalisms such as the one developed by \citet{Canuto1991} or \citet{Canuto1996} which assumes a Kolmogorov spectrum for the turbulent convective flow.

    A schematic flowchart of steps followed by \cesampp during the computation of a time step is represented in Fig. \ref{fig:cest_time_step}. At a given time step, \cesampp iterates following a Newton-Raphson scheme to find the solution to the structure equations. More precisely, using the stellar structure determined at the previous time step, \cesampp starts by finding the boundaries of the convective zones (CZ), then it evolves the chemical composition to the current time-step, and finally solves the stellar structure for that time. After each iteration, numerical convergence is assessed. If it is satisfied, the solution is accepted and new time step is computed. In general, this process is repeated several times to find an acceptable solution.

    \subsubsection{Scheme for entropy-calibration}

    As mentioned in the introduction, the value of $\amlt$ controls the adiabat of the convective envelope of the stellar model. This is easily seen in Fig. \ref{fig:alpha_s_match} where we present an average specific entropy profile extracted from a 3D atmosphere simulation, and entropy profiles computed from 1D stellar structures. Despite the fact that 3D and 1D models give very different entropy profiles in the superadiabatic layers, they can converge in deeper layers to the same adiabat, providing we find the correct value of $\amlt$.

    The procedure of entropy calibrations requires an intermediate step between the computation of the new chemical composition and the resolution of the structure equations. This intermediate step is part of the Newton-Raphson iteration described in \ref{fig:cest_time_step}. During that step, we compute a preliminary version of the entropy $s_{\rm ad,1}$ and the mean molecular weight (the reason is discussed in Sect. \ref{subsection:correction}) at the bottom of the convective envelope. These quantities are computed in an approximate stellar structure obtained with a given $\alpha_1 \equiv \alpha$, either from a previous time step, or from a previous structure iteration. With the values of $\teff$, $\log g$ and $\feh$ for this intermediate model, we evaluate the chosen functional representation of $\sad$ that provides the entropy $s_{\rm presc, 1}$ of the adiabat that this model should have.

    Then, the value of $\alpha_1$ is modified by a small amount: $\alpha_2 = \alpha_1 + \delta \alpha$ (currently, $\delta \alpha = 0.001 \alpha_1$). With $\alpha_2$, we call again the routine that solves for the stellar structure. Again, we compute $s_{\rm ad,2}$ at the bottom of CZ, and the value $s_{\rm presc, 2}$ that it must take according to the chosen functional. We want to find the correction $\Delta \alpha$ applied to $\alpha_1$, such that at next iteration we have:
    \begin{equation}
        \delta s_1 + \frac{\delta s_2 - \delta s_1}{\delta \alpha} \Delta \alpha = 0,
    \end{equation}
    with $\delta s_i = (s_{\rm presc,i} - s_{\rm ad,i})$, $i=1,2$. Then, $\Delta \alpha$ is:
    \begin{equation}
        \Delta \alpha = -\frac{\delta s_1}{\delta s_2 - \delta s_1}\delta \alpha.
    \end{equation}

    In the \cesampp input file, the fitting function proposed respectively by \citet{Ludwig1999}, \citet{Magic2013}, and \citet{Tanner2016} can be set through the option \texttt{nom\char`_alpha} respectively equal to \texttt{'entropy\char`_ludwig99'}, \texttt{'entropy\char`_magic13'}, and \texttt{'entropy\char`_tanner16'}.

    \begin{figure*}[t]
        \begin{center}
            \includegraphics[draft=False]{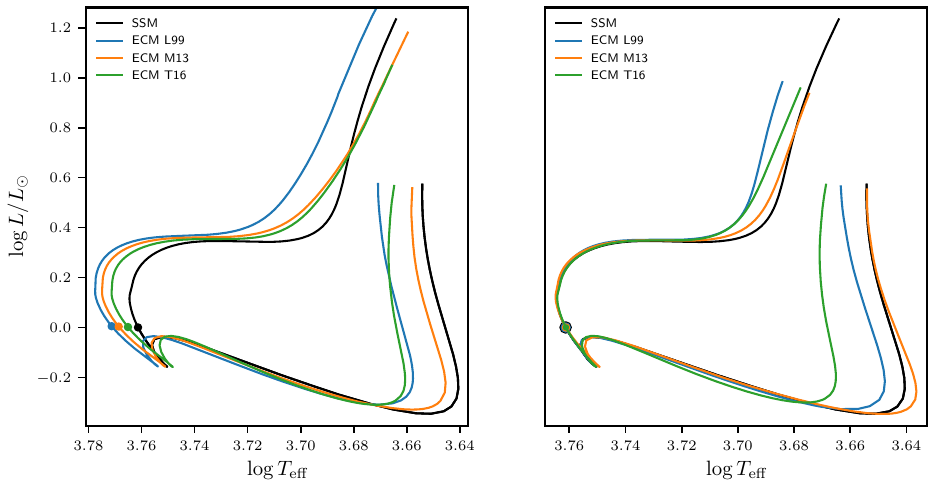}
            \caption{Left panel: HR diagram of a standard solar model (black line) and associated entropy-calibrated models with the uncorrected prescriptions L99 (blue), M13 (orange) and T16 (green) taken 'as is'. Dots represents the location of each model at solar age. Right panel: HR diagram of a standard solar model (black line) and associated entropy-calibrated models with the prescriptions L99 (blue), M13 (orange) and T16 (green) corrected using Eq. \eqref{eq:correc_final}. Dots represent  the location of each model at solar age.}
            \label{fig:HR_sun_ecm}
        \end{center}
    \end{figure*}

    \subsubsection{Control of precision and performance}

    The agreement between $\sad$ and $\spresc$ is controlled before the end of a time step. If the absolute value of the difference between the two exceeds $10^5~\mbox{erg}\,\mbox{g}^{-1}\,\mbox{K}^{-1}$, the solution is rejected and the computation of the current evolution step is started over again with a time step divided by two. This threshold was found by trial and error, and can be adjusted by the user. To give a comparison, in the \stagger grid, the total variation of the entropy of the adiabat spans $1.6903\cdot\sunit$. Therefore, the criterion is almost always met at the first try and the time step rarely needs to be decreased.

    In terms of performance, the use of entropy calibration significantly increases the computation time (by a factor 2 to 3). This is due to the fact that in order to evaluate the effect of a change of $\alpha$ on $\sad$, one has to recompute the entire stellar structure with $\alpha + \delta \alpha$. YREC proceeds differently: the envelope and the rest of the star can be computed separately and this can provide the new adiabat $s_{\rm ad, 2}$ much faster. In \cesampp, the interior and the envelope are solved together and the calculation cannot be decoupled.

    \section{Amendments to the entropy prescription}
    \label{section:sec4}

    \subsection{A first naive implementation}
    \label{subsection:naive}

    Once the entropy calibration procedure has been implemented into \cesampp, the next question is how the method impacts a standard\footnote{In the following, "standard" means that $\amlt$ stays fixed along evolution.} evolutionary track. To do so, we first calibrated a standard solar model (SSM) with a chemical composition following the determination of GS98 which is close  to the one used in the 2D atmosphere grid  which the L99 function was fitted on. The opacity and EoS are interpolated from the OPAL5 tables, convection is treated in the MLT framework, gravitational settling is modelled using the \citet{Michaud1993} approximation and no rotation is assumed. The atmosphere is reproduced using \corr{the reduced relation, known as the Hopf function, $q(\tau)$}:
    \begin{equation}
        T(\tau) = \teff \left[ \frac{3}{4}\left(q(\tau) + \tau\right)\right] ^{1/4}.
    \end{equation}
    The Eddington $T(\tau)$ relation uses a constant $q(\tau) = 2/3$. Other choices are discusses in Sect. \ref{subsubsection:ttau}.

    This SSM is calibrated by adjusting the initial helium mass fraction $Y_0$, initial metal to hydrogen ratio $(Z/X)_0$ content and the parameter $\alpha_{\rm MLT}$ to match, at solar age, the luminosity $1\,L_\odot$ \citep[$L_\odot = 3.828\times10^{33}\lunit$;][]{IAU2015}, the effective temperature $5772\,\K$, and the current surface $(Z/X)_{\rm s}$ of the Sun with a maximum  error of respectively $10^{-5}$ and $1\,\K$, $10^{-5}$. These parameters are adjusted using a Levenberg-Marquardt algorithm implemented in OSM (Optimal Stellar Model\footnote{\url{https://pypi.org/project/osm/}, developed in Python 2.7 by R. Samadi, adapted to Python 3 for the present work}; for detailed applications, see \citealt{Castro2021}). \corr{This package is currently interfaced with the \cesampp code and finds the model that matches the better a set of observational constrains (in that sense it is called and "optimal" model). Best values found for adjustable parameters of the model} are reproduced in Table \ref{table:optimal_sun}. Once this optimal model has been  obtained, we recompute the evolution by keeping everything identical, except that this time, the value of $\amlt$ is adjusted at each time step during its evolution to match the entropy prescribed by either L99, M13 or T16. These models are called entropy-calibrated models (ECMs).

    The resulting evolutionary tracks are represented in Fig. \ref{fig:HR_sun_ecm}, left panel. As can be seen, the four models follow very different paths. During the PMS, the locations of the respective Hayashi tracks are very far apart, while all models follow quite a similar Henyey track \citep{Henyey1955}. At solar age, the shift in effective temperature can reach $\simeq130\,\K$, with respect to the SSM. Models computed with M13 and T16 lead to similar results for evolved phases after the Terminal Age Main Sequence (TAMS), but they significantly differ from the one obtained with L99.

    ECMs are different from SSM, but are they better? The discrepancies obtained at solar age are problematic: the Sun must have a definite entropy for its adiabat, however, all prescriptions give very different values for the same set of input values. With this first implementation, it would seem that entropy-calibration is not very robust for stellar models. Moreover, the shift observed on the RGB between the track computed with L99 on the one hand, and the tracks computed with M13 and T16 on the other hand may reveal large differences between the grids on which these functions are adjusted. These issues stress the fact that the prescriptions should not be used at face value as given in the original papers but at least, corrected.

    \subsection{On the proper use of entropy prescriptions}
    \label{subsection:correction}

    In their works, \citet{Spada2018} and \citet{Spada2019} used the entropy given by T16's mathematical formulation corrected by an offset $\ds$. Indeed, the entropy is formally defined up to a constant. This constant can differ from one code to another because different EoS tables were used to generate either the atmosphere or the evolution models. Actually, entropies used in the \stagger-grid to calibrate the coefficients in M13 or T16's functions were already defined using a constant offset to make possible their comparison to the ones in L99 models (see \citealt{Magic2013}, p. 9).

    To compute the offset $\ds$ that should be added to $\spresc$, one needs to first compute a model with a standard, i.e. fixed $\alpha$, and then compare its adiabat to the one predicted by the chosen prescription. Of course, the standard model should use the same physics, and especially the same \eos, as the entropy calibrated model that will be computed.

    The second effect that needs to be taken into account is the entropy variation due to a change of chemical composition. Strictly speaking, to conduct entropy calibration, we should always consider models with a chemical composition identical to the one used to fit the coefficients of the entropy representation. However, the chemical composition changes through the star's life and this should be accounted for. In an ideal gas, neglecting the radiation pressure and assuming a polytropic \eos, one can write the entropy as \citep[e.g.][]{Ireland2018}:
    \begin{equation}
        s \simeq s_0 + \frac{N_{\rm A} k_{\rm B}}{\mu} \ln \left(\frac{T^{1/(\gamma - 1)}}{\rho} \right),
        \label{eq:s_def}
    \end{equation}
    where $s_0$ is a constant, $\mu$ is the mean molecular weight, $T$ the temperature, $\rho$ the density, and $\gamma$ the adiabatic exponent. A change of chemical composition has its main impact through the mean molecular weight. Because the term $s_0$ in Eq. \eqref{eq:s_def} is small compared to the second term\footnote{This has been verified using data of the 3D atmosphere models grid CIFIST reproduced in Table \ref{table:cifist0}.}, the entropy of a given atmosphere model with a mean molecular weight changing from $\mu_{\rm RHD}$ to $\mu$ can be approximated as
    \begin{equation}
        s \simeq \spresc f_\mu,~\textrm{with}~f_\mu = \frac{\mu_{\rm RHD}}{\mu}.
        \label{eq:fmu}
    \end{equation}
    This relation is valid only for a small change of $\mu$.

    This correcting factor was used in \citet{Spada2021} to account for significant chemical changes in evolved stellar phases but the offset correction $\ds$ was not used. We recommend however to use both corrections. So, the next issue is to know what form the correction should take. Coming back to the computation of the offset $\ds$, if the standard 1D model and the atmosphere model grid do not share the same chemical composition, then the offset simply computed as $\ds = \sad - \spresc$ does not only depends on a different choice of constant, but also on the different chemical mixtures. Instead, the offset should be adjusted as follows
    \begin{equation}
        \ds^\presc = \sad^{\rm std} - \spresc(\teff^{\rm std}, \logg^{\rm std}, \feh^{\rm std}) f_\mu^{\rm std},
    \end{equation}
    where \corr{"std" stands for quantities of the standard model}, $f_\mu^{\rm std}$ from \eqref{eq:fmu} accounts for the change in mean molecular weight, \corr{from the entropy prescription (based on 3D RHD simulations) to the 1D \corr{standard} model we want to apply it to.} Then, the corrected entropy takes the form
    \begin{equation}
        \spresc^{\rm corrected} = \spresc f_\mu - \ds^\presc.
        \label{eq:correc_final}
    \end{equation}
    This correction is applied every time a prescription is evaluated because $f_\mu$ changes in the course of evolution, e.g. as a result of gravitational settling or dredge up.

    \begin{table}
        \caption[Entropy offsets]{Entropy offset $\ds~[\sunit]$ computed with the SSM  model and for various prescriptions. The coefficients used in the prescriptions can either be taken from the original papers, or re-calibrated on the CIFIST grid (see appendix \ref{app:coeff}; \citealt{Ludwig2009}).}
        \label{table:ds_sun}
        \centering
        \begin{tabular}{L{0.5\linewidth} c c c }
            \hline\hline
                         & \multicolumn{3}{c}{Prescription} \\
                         & L99      & M13     & T16         \\
            \hline
            with original coefficients:    & $0.011$  & $0.042$ & $ 0.021$    \\[4pt]
            with coefficients recalibrated on the CIFIST grid:     & $0.008$  & $0.010$  & $-0.001$   \\
            \hline
        \end{tabular}
    \end{table}

    \begin{figure*}[t]
    \begin{center}
        \includegraphics[width=0.82\textwidth,draft=False]{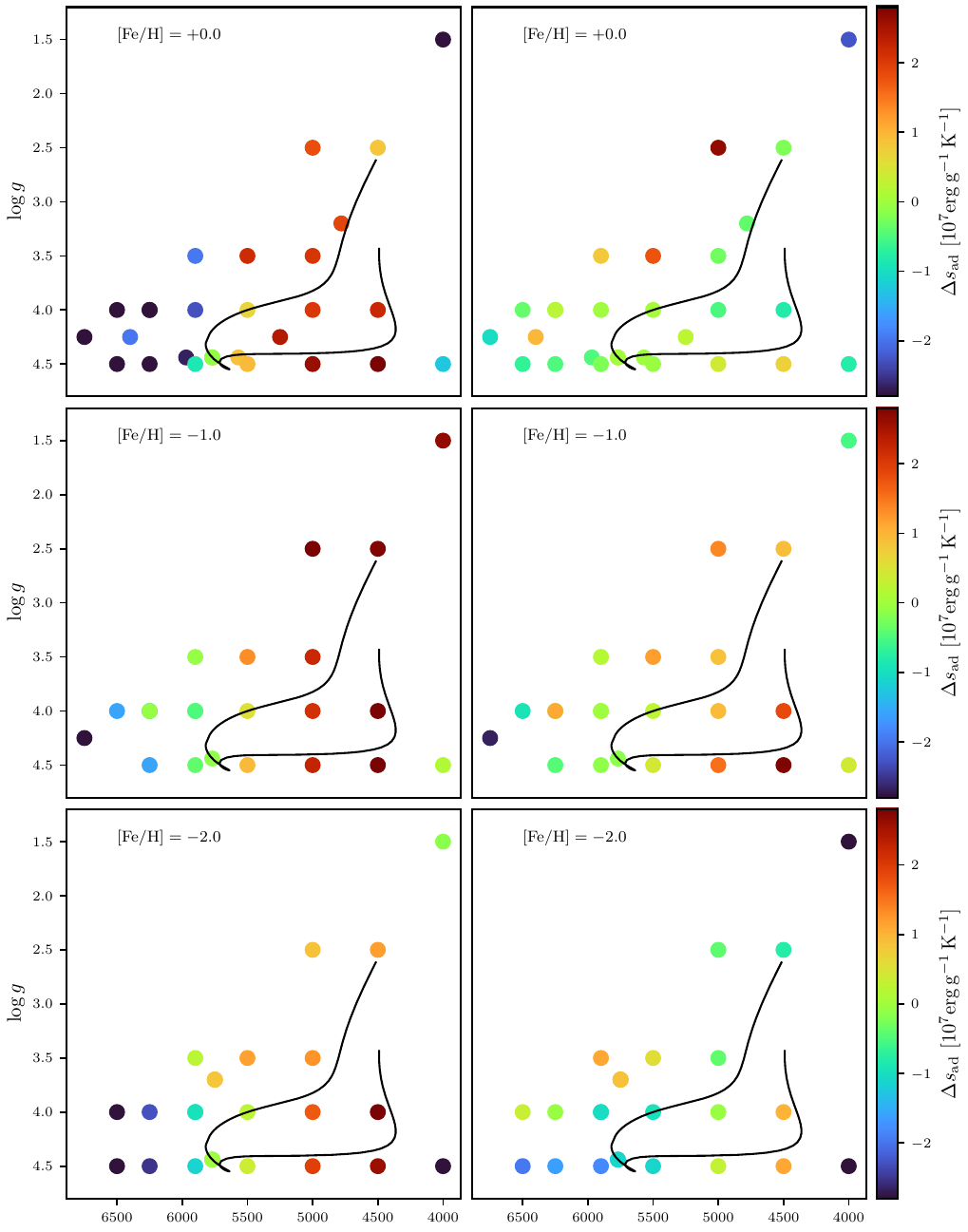}
        \caption{Specific entropy differences in the adiabat $\Delta s^\presc \equiv s_{\rm RHD} - s_{\rm presc}$ between values provided by the CIFIST grid \citep{Ludwig2009}, and values predicted by the M13 prescription, Eq. \eqref{eq:s_mag}. Left column: orignal coefficients from \citealt{Magic2013} are used. Here, entropy from the CIFIST grid has been corrected by an offset of $0.04794\cdot\sunit$ corresponding to the entropy difference between the solar atmosphere models in the CIFIST and \stagger grids. Right columns: Same as left column but the coefficients involved in the M13 mathematical form have been re-calibrated with the CIFIST entropies. Rows: Each row displays models with a given metallicity: $\feh = +0.0$ (1\st row), $-1.0$ (2\nd row) and $-2.0$ (3\rd row).}
        \label{fig:error_magic}
    \end{center}
\end{figure*}

In order to see the impact of this correction on the evolutionary tracks, we repeated the experiment of Sect. \ref{subsection:naive} but, this time, the ECMs follow the corrected prescriptions. To compute the entropy offset of each entropy representation, we used SSM as our standard model. The values are given in the 1\st line of Table \ref{table:ds_sun}. The evolutionary tracks of SSM and the new ECMs are presented in Fig. \ref{fig:HR_sun_ecm}. Now, we find that all four tracks intersect at solar age. However differences remain during the Hayashi tracks and on the RGB. The Hayashi track is a very short period of time ($\simeq 6~\mbox{Myrs}$ for the Sun), and its modeling is subject to a lot of uncertainties \citep[e.g.][]{Zwintz2022}. The differences seen in this phase do not propagate to later evolutionary phases and do not affect the validity of the method. The differences for PMS and RGB models have two sources. First of all, these two regions correspond to locations where the three entropy functional representations have the higher inaccuracies, i.e. largest errors between fitting function and entropy of the adiabat coming from 3D models. Moreover, the entropy of the adiabat does not vary linearly with $\teff$ and $\logg$. The entropy fitting functions of Eqs. \eqref{eq:s_lud},\eqref{eq:s_mag} or \eqref{eq:s_tan}, together with coefficients given in appendix \ref{app:coeff}, show that the entropy of the adiabat increases towards higher $\teff$ and lower $\logg$, but with decreasing slopes in the same direction. Therefore, a given uncertainty on the entropy coming from the fitting procedure, will imply a smaller change of location when the model is on the RGB region than when it is in the PMS. This explains why the discrepancies in the RGB region are smaller than around the Hayashi track. Still, some work remains to be done to improve the method in these evolutionary phases, e.g. by improving the analytic prescriptions in these regions of the Kiel diagram.

\subsection{Recommendations on the use of prescriptions}

In addition to the a posteriori correction of the prescribed entropy values, the 3D RHD simulation basis for the three formulations (L99, M13 and T16, that is a re-fit of M13) are not equivalent. The models used to calibrate the L99 law have been computed in 2D, therefore they may not be   reliable enough  to reproduce the entropy stratification of 3D models. Furthermore, this relation was obtained only for one (solar) metallicity. Therefore, the correction $f_\mu$ may not be valid because the difference between $\mu_{\rm RHD}$ and $\mu$ from the entropy calibrated model may not be close enough to zero. For these reasons, we recommend not to use the L99 prescription (with any set of coefficients), except when computing non diffusive models, with a chemical composition similar to GS98. L99 is therefore not used in the remainder of this work.

Regarding the functional forms suggested by M13 and T16, it appears that they have not been originally calibrated against the 'true' entropies  of the adiabat of the \stagger grid (that are inputs of the models), but against $s_{\rm bot}$ the average entropy at the bottom of the simulation domain ($\srhd$ values are not publicly available). These simulations assume that at the bottom of the boxes, the adiabat is reached and $s_{\rm bot} = \srhd$. However, this is only true for some models (see Table \ref{table:cifist0}) and it introduces significant disagreements between $\srhd$ and the value given by the prescription. On the contrary, in \cobold models of the CIFIST grid, the true entropy of the adiabat \corr{is known exactly because it} is an input of the simulation and corresponds to the entropy of the inflow at the bottom boundary of the simulation domain. \corr{This also mean that the value of $\srhd$ is not flawed by any boundary effect, which is not the case for the value of $s_{\rm bot}$.}

In order to perform a comparison of the various entropy representations (see Sect.~\ref{section:sec2}) and to improve the quality of the entropy calibration, we re-calibrated the coefficients on a set of entropies of the adiabats obtained from 102 models of the CIFIST grid \citep{Ludwig2009} computed with the \cobold code \citep{Freytag2002,Freytag2012,Wedemeyer2004}. Here, we use the true entropy of the adiabats, $\srhd$, not the one computed at the bottom of the simulated box. These data are available in appendix \ref{app:grid} for models with solar metallicity\footnote{Data for all metallicities are available at CDS (\url{http://cds.u-strasbg.fr}).} and the coefficients of the functional forms are available in appendix \ref{app:coeff}. Figure~\ref{fig:error_magic} illustrates the impact of re-calibrating the M13 function, defined in Eq. \eqref{eq:s_mag}. The left column shows $\Delta s^{\rm M13, o} \equiv \srhd^{\rm CIFIST} - \spresc^{\rm M13, o}$, i.e. the entropy difference between the true entropy of the adiabat in the CIFIST grid and the original M13 formulation (original in the sense that coefficients where taken from \citealt{Magic2013}). Note here that a constant has been added to the original M13 relation to compensate for the zero point offset between the entropy of the solar models in the CIFIST and \stagger grids, as stated in Fig. \ref{fig:error_magic} caption. The right column in Fig.~\ref{fig:error_magic} shows the differences $\Delta s^{\rm M13, r} \equiv \srhd^{\rm CIFIST} - \spresc^{\rm M13, r}$, in which $\spresc^{\rm M13, r}$ is obtained after the M13 formulation has been re-calibrated in the CIFIST grid.

\begin{figure}[t]
        \begin{center}
            \includegraphics[draft=False]{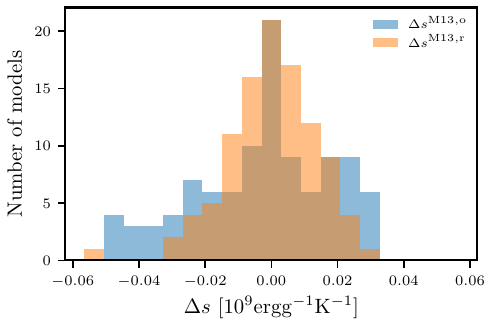}
            \caption{Histogram showing the distribution of the entropy errors $\Delta s^\presc$ for the  M13 prescription. The data plotted are the same as shown in Fig. \ref{fig:error_magic}, for all metallicities. In blue is shown the $\Delta s^{\rm M13, o}$ distribution corresponding to differences $\srhd^{\rm CIFIST} - \spresc^{\rm M13, o}$, where $\spresc^{\rm M13, o}$ are the entropies predicted by the M13 prescription with original coefficients (taken from \citealt{Magic2013}). The $\Delta s^{\rm M13, r}$ distribution is shown in orange, corresponding to $\srhd^{\rm CIFIST} - \spresc^{\rm M13, r}$, where this last term is the entropies obtained from the M13 prescription re-calibrated on the CIFIST grid (therefore on the true entropies of the adiabats).}
            \label{fig:error_magic_histo}
        \end{center}
    \end{figure}

The distribution of $\Delta s^\presc$ for CIFIST models is also represented as a histogram in Fig.\ref{fig:error_magic_histo}. We see that this new calibration reduces the range of entropy differences (from $\simeq 0.05$ to $\simeq 0.03\cdot \sunit$) but the accuracy of M13 remains similar for the two grids. However, variations of $\Delta s^\presc$ across the Kiel diagram do not behave the same. In Fig. \ref{fig:error_magic}, left column, the entropy difference rises with decreasing $\teff$, but in the right column, the larger differences are localized in the PMS region and around the RGB of stars more massive than the Sun. Those differences may arise from two effects: \textit{(1)} the offset we added does not fully correct the discrepancies between the different EoS and reference solar composition (i.e., these differences do not induce a constant change across the Kiel diagram); \textit{(2)} in Fig. \ref{fig:error_magic}, left column, coefficients of Eq. \eqref{eq:s_mag} are calibrated on $\sbot$ instead of $\spresc$.

\begin{figure}[t]
        \begin{center}
            \includegraphics[draft=False]{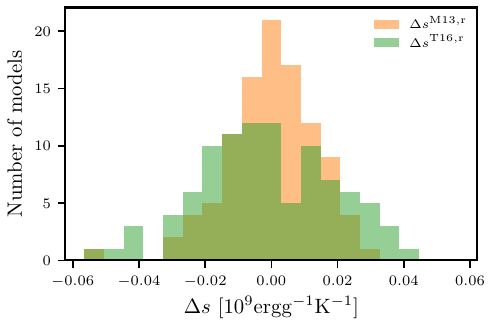}
            \caption{Histogram showing the distribution of the entropy errors between $\srhd$ and $\spresc$ for M13 and T16 entropy representations. Both functions have been re-calibrated on entropies from the CIFIST grid.}
            \label{fig:error_presc_histo}
        \end{center}
    \end{figure}

    The distributions of $\Delta s^\presc$ for the re-calibrated M13 and T16 formulations (Eq. \eqref{eq:s_tan}) are represented in Fig.~\ref{fig:error_presc_histo}. Both prescriptions hardly exceed an absolute error of $0.05\cdot \sunit$, which is less than $3\%$ of the extremal variation of entropy across the CIFIST grid. This agreement is very satisfying. In addition, we note that the distribution of $\Delta s^{\rm M13, r}$ is much more peaked around 0.0 than the one of $\Delta s^{\rm T16, r}$. This comparison advocates for generalizing the use of the M13 prescription in entropy-calibrated modeling. From this point onwards, we only use the M13 prescription of adiabatic entropy.

    \section{Discussion}
    \label{section:sec5}

    \subsection{Possible sources of uncertainties}

    The accuracy of entropy calibration modeling can be affected by several factors. In the following section, we review them and estimate their impact   on the determination of a star's characteristics. One source of uncertainty is the already mentioned gap between $\srhd$ and the corresponding value obtained from prescriptions. Accuracy can also be impacted by the methods used to determine the offset $\ds$, and the $\sad$ and $\mu$ values at the bottom of the convective envelope of the stellar evolution model. Then, we should also examine what happens when a prescription is evaluated outside of its domain of definition. Finally, it could be expected that a different choice of $T(\tau)$ relation would change the location of an evolutionary track on the HRD or Kiel diagrams. These points are investigated in the following subsections.

    \subsubsection{Determination of $\mu$ and $\sad$ in 1D models}

    Several factors can affect the accuracy of entropy-calibration modeling. The first one comes from the way we determine the mean molecular weight $\mu$ and the entropy of the adiabat in the 1D model.

    The mean molecular weight depends on the relative abundance and on the ionization rates of each element. Oftentimes, stellar atmosphere models do not provide the variables needed to compute $\mu$ accurately. However, we can assume that, at the bottom of the convective zone, the material is fully ionized. Then, $\mu$ can be approximated as $\mu^{-1} = 2X + 3Y/4 + Z/2$ where $X$, $Y$ an $Z$ are respectively the hydrogen, helium and metal abundance. The error induced by this approximation is below $1\%$ for stars with $M < 1.5 M_\odot$. We could relax this approximation by assuming that the ionization rates are the same in the evolutionary and in the atmosphere models. This would allow to consider each element individually, instead of a dependence in $X$, $Y$ and $Z$. However, for this to work, it would require to consider exactly the same chemical elements in both simulations, which would be too much constraints from a modelling point of view. In any case, this would modify the value of $\mu$ by an almost constant factor, that would be taken into account in $f_\mu$, defined in Eq. \eqref{eq:fmu}.

    Concerning the determination of the entropy of the adiabat in the 1D model, we compared three methods. In the first one, we define $\sad\equiv\sad^{\rm bot}$ as the value of the entropy exactly at the transition between the radiative zone and the convective envelope. Another method would be to define $\sad\equiv\sad^{\rm av}$ as the average of the entropy in a sub-region of the adiabatic region. This second method requires to interpolate in the EoS table for each layer  in the sub-region, which is time consuming. To improve its efficiency, we rather introduce a third definition where $\sad\equiv\sad^{\rm av10}$ is the average of the entropy computed for only every 10 layers in the sub-region.

    \begin{figure}[t]
        \begin{center}
            \includegraphics[width=\linewidth,draft=False]{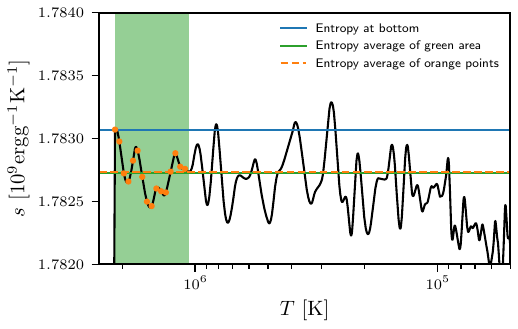}
            \caption{Entropy as a function of temperature (black line) in a sub-region of the convective envelope of SSM. The value of entropy at the limit between radiative and convective zone is represented as a blue line. The average of entropy in the green shaded area is represented as a green line and the average of entropy at orange points is represented as a dashed orange line. The green area has a width of $10\%$ of the total width of the CZ (roughly $\nicefrac{2}{3}$ of the mass of the CZ) and the orange points are located each 10 layers in the green area.}
            \label{fig:sbot}
        \end{center}
    \end{figure}

    Fig. \ref{fig:sbot} represents the entropy profile in the adiabatic region of SSM. We stress that the figure is extremely zoomed-in and what seems to be large fluctuations is actually nearly constant. Starting from $T\simeq 10^5~[\K]$ to the right, we see a decrease of entropy, corresponding to the beginning of the superadiabatic region. The value of $\sad^{\rm bot}$ is represented as the blue line. The green shaded area corresponds to the sub-regions in which $\sad^{\rm av}$ and $\sad^{\rm av10}$ are defined. The green region represents $10\%$ of the radial extension of the convective envelope and $\simeq 66\%$ of its total mass. This extent is chosen arbitrarily but ensures that, for every model in the domain of definition of the CIFIST grid, we only enclose a sub-region of the adiabatic region. Clearly, $\sad^{\rm av}$ and $\sad^{\rm av10}$ are nearly equal. So, does the use of $\sad^{\rm av10}$ instead of $\sad^{\rm bot}$ give significantly different results ?

\begin{figure}[t]
        \begin{center}
            \includegraphics[width=\linewidth,draft=False]{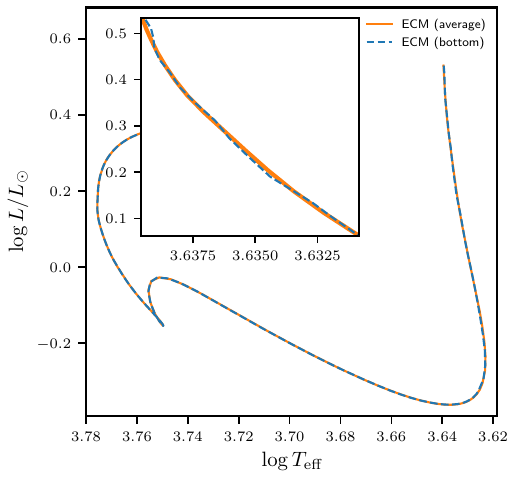}
            \caption{HR diagram of the same entropy-calibrated model, using the M13 prescription, but different methods for evaluating the entropy of the adiabat in the 1D structure model: $\sad^{\rm av10}$ in orange and $\sad^{\rm bot}$ in blue. The subplot shows a zoom of the Hayashi track.}
            \label{fig:HR_s_determ}
        \end{center}
    \end{figure}

    We implemented both definitions in \cesampp and computed two evolutionary tracks using the M13 prescription. Tracks are shown in a HR diagram in Fig. \ref{fig:HR_s_determ}. In terms of global quantities, the two definitions give similar results. However, as stressed by the subplot showing a zoom of the Hayashi track, using $\sad^{\rm bot}$ gives a more irregular path than $\sad^{\rm av10}$, and using $\sad^{\rm av10}$ actually greatly increases the numerical stability of the method.

    \subsubsection{Constancy of the entropy offset $\ds$}

    The way we computed the offset $\ds$ immediately brings the concern that it may vary across HR diagram. As said before, the offset $\ds$ corrects for a systematic difference in the entropy of the 3D atmosphere grid and the one computed in the stellar evolution model. This offset is the sum of the entropy zero-point for that EoS, and an ad-hoc constant for it to match the L99 entropy. The offset should be estimated with a standard solar model (standard in the sense that $\amlt$ stays fixed along evolution), which provides us with the most accurate constraint one can obtain on the entropy of the adiabat.

    \begin{figure*}[t]
        \begin{center}
            \includegraphics[width=\linewidth,draft=False]{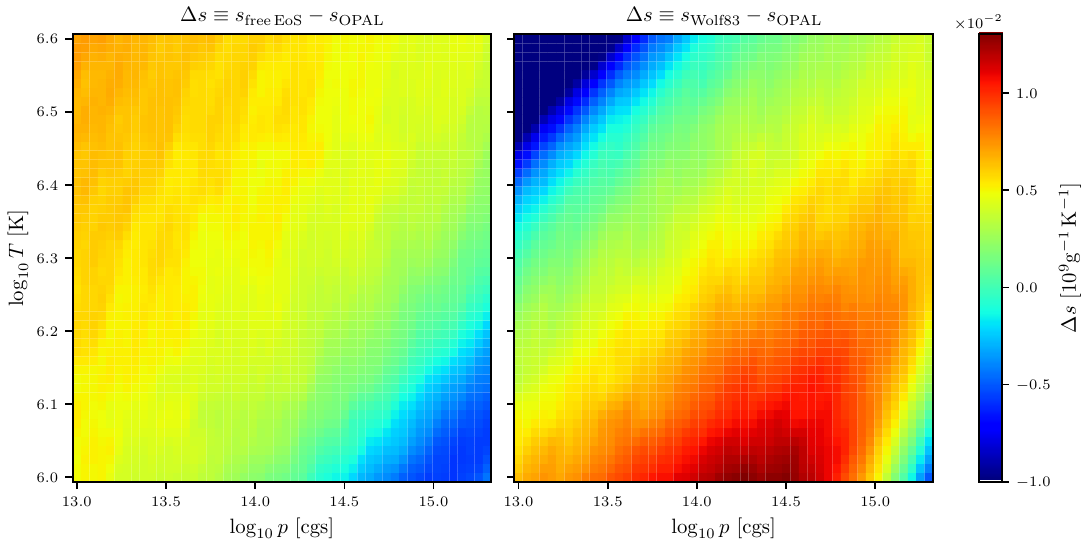}
            \caption{Entropy differences between two EoS, in the $\log_{10} p - \log_{10} T$ plane. The colour codes for the entropy difference. Left panel: entropy differences between free EOS and OPAL5. Right panel: entropy differences between EoS of \citet{Wolf1983} and OPAL5.}\label{fig:ds_EOS_2}
        \end{center}
    \end{figure*}

    In order to verify that $\ds$ is almost constant over the HR diagram, we computed entropy for three different EoS (numerical or analytical), for a given chemical composition, and conditions of pressure and temperature that cover the ones found at the bottom of convective envelope of stellar models with a mass between $0.8$ to $1.3~M_\odot$. The three EoS are the OPAL5 equation of state, the simple EoS proposed by \citet[hereafter W83\footnote{Notice the typo in the summation part of their Eq. (16): one should not read $m_{\rm e}$, the mass of an electron, but $m_i$ the mass of the considered element.}]{Wolf1983} and FreeEOS \citep{Irwin2012}. OPAL5 is the EoS table used to compute \cesampp models. The simple analytic EoS proposed by W83 is the one used by \cobold. Here, we implemented a simplified version which accounts only for ${\rm H}^+$, ${\rm He}^+$, ${\rm He}^{++}$, ${\rm e}^-$ and a representative metal (\textit{case f.} in W83). FreeEOS is also an analytic EoS but much more detailed. We followed FreeEOS's recommendation and adopted the version suitable for stellar interiors that also reproduces the MHD equation of state \citep{Mihalas1988} used by the \stagger code. This version includes 20 elements with 295 ionization states. The entropy differences between FreeEOS (resp. W83) and OPAL5 are presented in a $\log_{10} p - \log_{10} T$ plane in left (resp. right) panel of Fig. \ref{fig:ds_EOS_2}. Extremal variation in this interval is of the order of $10^7~\erg\,\mbox{g}^{-1}\,\mbox{K}^{-1}$. This is negligible compared to, e.g., the extremal variation of entropy across the CIFIST grid ($1.69\cdot\sunit$). In addition, these extremal variations are reduced if we exclude top left (resp. bottom right) corner that corresponds to regions of low $p$ -- high $T$ (resp. high $p$ -- low $T$) which are never found at the bottom of the convective envelope of our models. Therefore, the assumption that offset $\ds$ is constant over the HR diagram holds to good accuracy.

    \subsubsection{Impact of the $T(\tau)$ relation}
    \label{subsubsection:ttau}

    \begin{figure*}[t]
        \begin{center}
            \includegraphics[width=\linewidth,draft=False]{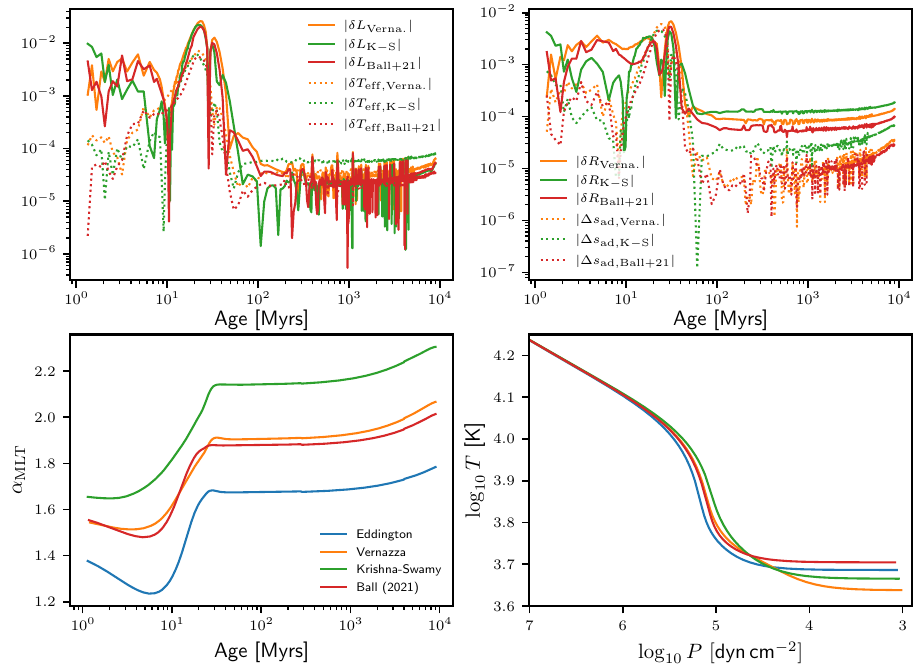}
            \caption{Top left panel: Comparison of luminosity (solid lines) and effective temperature (dashed lines), for four models computed with different choices of $T(\tau)$ relation. The displayed quantity is the relative difference of $L$ or $\teff$, with respect to model computed with the Eddington $T(\tau)$ relation: $\delta X \equiv (X_{T(\tau)} - X_{\rm Edding.}) / X_{\rm Edding.}$. Top right panel: same as top left panel but, in the case where $X \equiv \sad$, we display the absolute difference. Indeed, $\sad$ being defined up to a constant, the relative difference would have no meaning. $\Delta s^\presc$ is in the units: $\sunit$. Bottom left panel: Evolution of $\amlt$ for different $T(\tau)$ relations: Eddington (blue), \citet{Vernazza1981} Model-C (orange), \citet{Krishna-Swamy1966} (green) and \citet{Ball2021} (red). Bottom right panel: Temperature as a function of the pressure for the same four final models, at solar age.}
            \label{fig:ttau}
        \end{center}
    \end{figure*}

    The previous attempts to prescribe a value for $\amlt$ have in common the weakness that the $\alpha$ prescription depends on the kind of $T(\tau)$ relation chosen to reproduce the atmosphere. This issue disappears with entropy calibration modeling, as shown in \citet{Spada2018} for the Sun and in \citet{Spada2021} for MS and RGB stars. The reason is that the entropy of the adiabat depends on the characteristics of the adiabatic layers, not on how the atmosphere is computed. The adiabat of the convective envelope of  a star  depends solely on the star's location in the Kiel diagram and its metallicity. If the $T(\tau)$ relation changes, the value of $\amlt$ will adapt, but the location on the Kiel diagram is the same. To verify that this property remains true with our procedure, we computed four identical solar entropy calibrated models except for the $T(\tau)$ relation that follows either an Eddington law, a fit of \citet{Vernazza1981}'s Model C of the quiet Sun, the \citet{Krishna-Swamy1966} relation, or a $T(\tau)$ relation recently proposed by \citet{Ball2021}.

    A change of $T(\tau)$ relation, at fixed $\amlt$ changes global quantities such as $\teff$. However, here, we are adjusting $\amlt$ to match a given entropy of the adiabat, therefore changing the $T(\tau)$ relation does not imply a change of location in the HR diagram. In the top panels of Fig.~\ref{fig:ttau} we represent the evolution with age, starting from PMS, of the relative $L$, $R$, $\teff$ and the absolute $\sad$ differences for models computed with various $T(\tau)$ relations, with respect to a standard model computed with the Eddington $T(\tau)$ relation. It clearly shows the invariance of such global quantities when a different $T(\tau)$ law is chosen. On the bottom left panel, we show $\amlt$ along evolution for the same models. Since, $\teff$, $\logg$ and the metallicity are the same from one model to the other, $\sad$ also remains constant. The structure of superadiabatic layers at final age (4~570 Myrs) is shown in the bottom panel of Fig. \ref{fig:ttau}. It stresses that $T(\tau)$ relations have an important impact on the modelling of the shallowest layers but the structures are all identical when deep enough in the star. The deepest layer shown in the figure has a pressure of $10^7~\punit$, which is well above the bottom of the convective envelope, located around a pressure of $10^{13}~\punit$. It stresses the fact that although the choice of $T(\tau)$ relation does not impact global properties of the model, this choice is still crucial for obtaining realistic models for asteroseismology.

    The examination of the bottom left panel of Fig. \ref{fig:ttau}, reveals a minimum in the $\amlt$ variation with age between 2 and 6 Myrs, when models are on their Hayashi track. It is not possible to relate this behaviour to previous calibrations of $\amlt$ \citep[e.g.][]{Ludwig1999,Trampedach2014,Magic2015}, except the calibration of \citet{Sonoi2019} for what they called "Eddington + MLT(BV)". This last work calibrated $\amlt$ by matching 1D CESTAM structure models to 3D CIFIST atmosphere models. CESTAM (and also its successor \cesampp) implements two versions of the MLT that assume either a linear temperature distribution in the convective bubble (that \citealt{Sonoi2019} called "BV"), or a distribution that \corr{obeys} a diffusion equation as suggested by \citet{Henyey1965}. The BV version of MLT is also the version used in this work, which explains the agreement.

    \subsubsection{Interpolation outside the prescription's domain of definition}
    \label{subsubsection:domain}

    It may happen that at some stage of the evolution, a model falls outside the domain of definition of the prescription, which is the domain covered by the CIFIST grid in this work. For instance, when starting from the PMS, a solar model spends its first 1 Myrs slightly outside this domain.

    Every time a given prescription is evaluated, \cesampp checks if the model (its $(\teff,\logg,\feh)$) is inside the convex hull of the set of 3D models. By default \cesampp only sends a warning (this behaviour can be changed by the user), and proceeds to evaluate the prescription. It must be noted that if prescriptions are evaluated outside of the domain, on the hot side, the resulting entropy can diverge because of the positive term in the exponential of Eqs. \eqref{eq:s_lud} to \eqref{eq:s_mag}). On the cool side, the exponential should remain small, and the entropy does not diverge (which does not mean its value is reliable). The only solution to circumvent this issue is to extend the range of definition of the prescriptions. This work is currently in progress.

    \subsection{Comparison with YREC: the \acen system}

    \begin{figure*}[t]
        \begin{center}
            \includegraphics[width=\linewidth,draft=False]{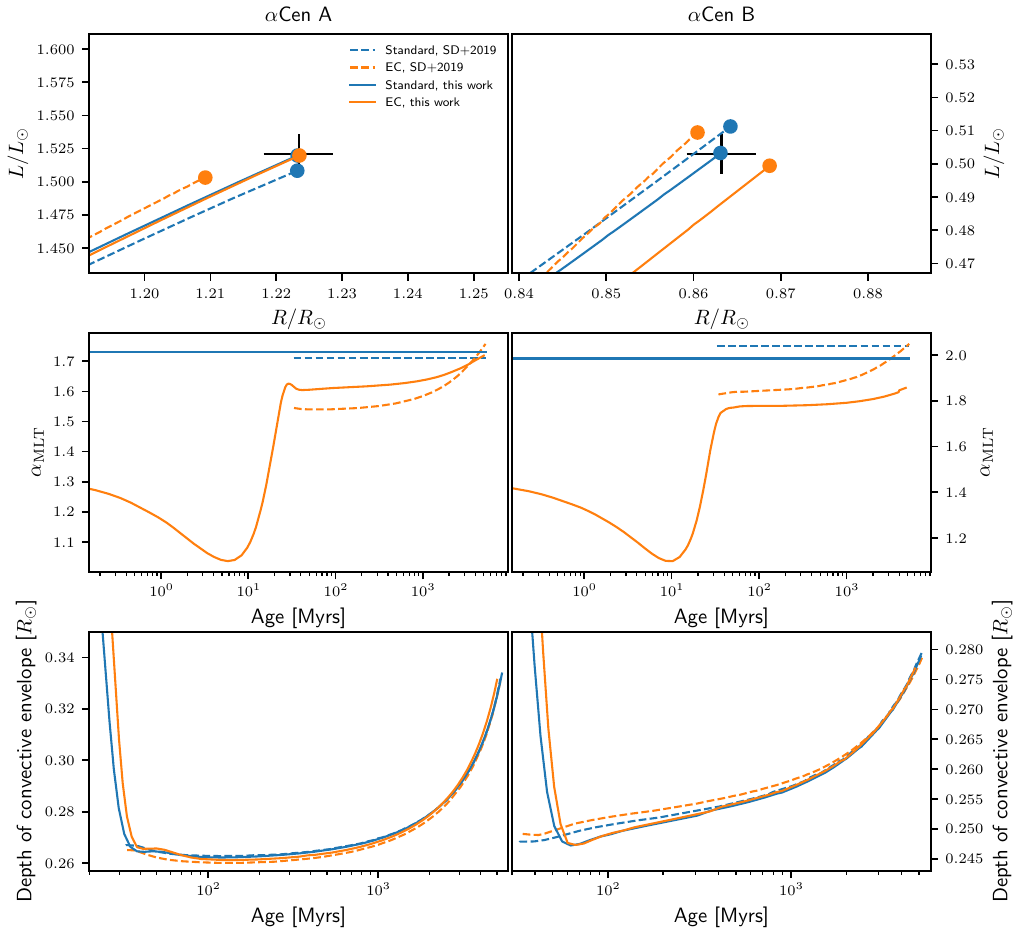}
            \caption{First row: Evolutionary tracks for \acen A (left panel) and \acen B (right panel) on the $R - L$ plane. Tracks computed in the present work are represented as solid lines, while the one from \citet{Spada2019} are drawn as dashed lines. Standard models are in blue and EC models in orange. Second row: Evolution of $\amlt$ as a function of age for the same models. Third row: Evolution of the convective envelope depth along evolution.}
            \label{fig:rl_alpha_dcz}
        \end{center}
    \end{figure*}

    \begin{table*}[t]
        \caption{Observable constraints and optimal parameters derived for the \acen system.}
        \label{table:optim_acen}
        \begin{center}
        \begin{tabular}{l c c c c c  }
            \hline
            Parameter               & Obs. constraints        & \multicolumn{2}{c}{Standard} & \multicolumn{2}{c}{Entropy-calib.} \\
                                    &                         &               &              &               &                    \\
                                    &                         & YREC (SD19)   & \cesampp     & YREC (SD19)   & \cesampp           \\
            \hline \\
            \multicolumn{6}{c}{\acen A}                                                                                     \\
            Mass ($M_\odot$ )       & $1.1055\pm0.004 $ (1)   & --            & --           & --            & --                 \\
            Radius ($R_\odot$ )     & $1.2234\pm0.0053$ (1)   & $1.2235$      & $1.2233$     & $1.2097$      & $1.2235$           \\
            Luminosity ($L_\odot$ ) & $1.521\pm0.015  $ (1)   & $1.5091$      & $1.5197$     & $1.5045$      & $1.5198$           \\
            $(Z/X)_{\rm s}$           & $0.039\pm0.006  $ (2,3) & $0.0315$      & $0.0327$     & $0.0347$      & $0.0361$           \\
            Age (Gyrs)              & --                      & $5.243$        & $5.35$       & $5.243$        & $5.00$           \\
            $\amlt$                 & --                      & $1.71$        & $1.732$      & $1.758^*$       & $1.719^*$            \\
            $Y_0$                   & --                      & $0.2785$      & $0.2781$     & $0.2785$      & $0.2898$           \\
            $Z_0$                   & --                      & $0.027$       & $0.0275$     & $0.027$       & $0.0298$           \\
            $\chi^2$                & --                      & $2.21$        & $1.12$       & $8.38$        & $0.24$             \\
                                    &                         &               &              &               &                    \\
            \multicolumn{6}{c}{\acen B}                                                                                     \\
            Mass ($M_\odot$ )       & $0.9373\pm0.003$ (1)    & --            & --           & --            & --                 \\
            Radius ($R_\odot$ )     & $0.8632\pm0.004$ (1)    & $0.8646$      & $0.8631$     & $0.8608$      & $0.8687$           \\
            Luminosity ($L_\odot$ ) & $0.503\pm0.006 $ (1)    & $0.5122$      & $0.5032$     & $0.5103$      & $0.4994$           \\
            $(Z/X)_{\rm s}$           & $0.039\pm0.006 $ (2,3)  & $0.0346$      & $0.0357$     & $0.0364$      & $0.0339$           \\
            Age (Myrs)              & --                      & $5.263$        & $5.11$       & $5.263$        & $4.75$           \\
            $\amlt$                 & --                      & $2.04$        & $1.985$      & $2.05^*$        & $1.86^*$             \\
            $Y_0$                   & --                      & $0.2668$      & $0.2666$     & $0.2668$      & $0.2640$           \\
            $Z_0$                   & --                      & $0.0273$      & $0.0278$     & $0.0273$      & $0.0263$           \\
            $\chi^2$                & --                      & $3.00$        & $0.30$       & $2.035$       & $2.99$             \\
            \hline
        \end{tabular}
        \end{center}
        Note. (1): \citet{Kervella2017}; (2): \citet{Thoul2003}; (3): \citet{PortodeMello2008}. ${}^*$: The values of $\amlt$ are the final values of the model, as for ECM, $\amlt$ is varying through evolution, as shown in Fig. \ref{fig:rl_alpha_dcz}, middle row.
    \end{table*}

    As a final test, we compared the results of modeling the \acen system computed with standard or entropy-calibrated modelling, to the models presented in \citet[][hereafter SD19]{Spada2019}. The microphysics of the models is kept as close  as possible to the one used in YREC models: MLT treatment of convection, Eddington atmosphere, element abundance ratios following the GS98 chemical composition and only gravitational settling is taken into account in the diffusive treatment. SD19 used the YREC code to obtain best fitting models of \acen A and B, using a MCMC method (more details are given in the SD19 paper). The use of a MCMC method allowed the authors to provide reliable uncertainties to their best estimated parameters. To speed up the process, evolution models were started at the ZAMS.

    In the present work, we found optimal \cesampp models for the \acen system using the OSM program. In our calibrations, we used the same observational constraints as the one used by SD19, summarized in Table \ref{table:optim_acen} and we adjusted the age of the model, its initial helium content $Y_0$, its initial $(Z/X)_{0}$ ratio and, only for the standard models, the $\amlt$, so that we match the observed radius, luminosity, and current $(Z/X)_{\rm s}$. In ECMs, of course, $\amlt$ is varying along evolution, following the change of entropy. The mass is not a free parameter: we used the value reported in Table \ref{table:optim_acen}. In parallel, using the same physical ingredients, we calibrated a solar model to be the standard model for the computation of entropy offset $\ds$. Using the  M13 prescription yields $\ds = 0.0264\cdot\sunit$ and we used this value to compute entropy-calibrated models by tuning the same parameters as before (except for $\amlt$). The Levenberg-Marquardt scheme implemented in OSM cannot yield reliable uncertainties and therefore, we give our best estimated parameters, without uncertainties. The squared-distance $\chi^2$ between the best \cesampp model and the observational constraints is computed with:
    \begin{equation}
        \chi^2 = \sum_i \left(\frac{X^i_{\rm obs} - X^i_{\rm model}}{\sigma_i}\right)^2,
    \end{equation}
    where $X^i_{\rm obs}$ are the observational constraints, $\sigma_i$ their uncertainties, and $X^i_{\rm model}$ are the values of the same parameters in the best model. The MCMC method used by SD19 explores a wide parameter space, and then requires the computation of more models than OSM. Therefore, we could afford to start our \cesampp models from the PMS.

    The optimal parameters obtained with YREC and \cesampp are presented in Table \ref{table:optim_acen}. Top panels of Fig. \ref{fig:rl_alpha_dcz} show the tracks of all models in the $\mbox{Radius}-\mbox{Luminosity}$ plane with the measured location of \acen A and B. Models taken from SD19 show slightly less good agreement with observational constraints than models computed for the present work. This is only a consequence of the optimization scheme: since the MCMC method explores all the parameter space, a very large number of models need to be computed, and obtaining a very precise solution is extremely time consuming. OSM requires to compute only a small number of models and the optimal solution is guided to a (possibly local) minimum.

    The differences obtained between YREC and \cesampp models can have many origins. First of all, the numerical schemes differ. The convective envelope is treated separately from the core in YREC, which allow it to tune $\alpha$ by recomputing only the envelope structure. This is not possible with \cesampp and we have to recompute the complete structure every time $\alpha$ is changed. Also, in the case of ECM models, different entropy prescriptions are used and the way of correcting them differs. SD19 also made use of the T16's function with the coefficients taken from the original paper while we used the M13's function with coefficients re-calibrated on the CIFIST grid. In addition, it must be stressed that the model of \acen B has its Hayashi track outside the domain of definition of M13, which was not the case for models in SD19, because their tracks started from the ZAMS. However, it represents a very short time of the life of \acen B and should not have significantly impacted its evolution. Finally, SD19 corrected the prescribed entropy using only the entropy offset while \cesampp also corrected for differences in the chemical composition. These discrepancies translate in quite different evolution of $\amlt$ between models with similar physics. However, the general trend remains similar: $\amlt$ increases when $\sad$ increases. The variations of entropy can easily be understood with the approximation that $\sad \propto \ln T^{\nicefrac{3}{2}}/\rho$. With the Virial theorem it comes that $T \propto R^{-1}$ and $\rho \propto R^{-3}$, $R$ being the total radius of the star. Therefore, during expansion phases, when $R$ increases, $\sad$ increases, and conversely during contraction phases.

    Bottom panels of Fig.~\ref{fig:rl_alpha_dcz} display the variation of the depth of the convective envelope $d_{\rm cz}$ for all models of the \acen system. Once again, it reveals a very good agreement between YREC and \cesampp models. Whereas $\amlt$ depends a lot on the physical ingredients and numerical methods used in the code and does not have a precise physical meaning, $d_{\rm cz}$ is a measurable quantity. Having such an agreement for $d_{\rm cz}$ between the two codes advocates for the reliability of the method. Moreover  the values of $d_{\rm cz}$ predicted for a standard and EC models are close, but not identical. For \acen B (the less massive), the two values are almost equal during the main sequence, while in the case of \acen A (the more massive), the ECM predicts a shallower CZ at the ZAMS and deeper at the present age of \acen A. If confirmed, such pattern should have repercussions on the transport of chemicals and angular momentum.

    \section{Conclusion}
    \label{section:sec6}

    We presented in this work a comprehensive description of the entropy-calibrated modelling: how to properly implement it and use it. The entropy-calibration consists in adjusting the value of the convection parameter $\alpha$ so that the entropy of the adiabat in a stellar structure model matches the one predicted by an entropy prescription.

    We studied and compared the accuracy of the three available prescriptions for the entropy of the adiabat suggested in \citet{Ludwig1999}, \citet{Magic2013} and \citet{Tanner2016}. Those functions are adjusted to fit the entropy of the adiabat in stellar atmosphere simulations, as a function of $\teff$, $\logg$ and $\feh$ in a grid of such simulations. Our study pointed out the fact that the M13 fit was based on the entropy, \corr{averaged over both the adiabatic up-flows and the cooled down-flows}, at the bottom boundary of their RHD simulations, instead of on the entropy of the asymptotic adiabat, and T16 was a re-fitting of those same, flawed entropies, introducing a bias on the prescribed entropy value. After the re-calibration of the entropy functional representations using the correct set of entropies of the adiabat obtained from the CIFIST grid \citep{Ludwig2009}, we showed that the functional form that most accurately reproduces the RHD results, and therefore the one we recommend to use, is the formulation proposed by \citet{Magic2013} with coefficients given in Table \ref{table:coeff_m13}. We make publicly available the quantities extracted from the CIFIST grid that have been used in this work. The coefficients of the fitting functions are also given in appendix \ref{app:coeff}, which extends the small list of entropy prescriptions available to the stellar modelling community.

    Not only the prescription must be carefully chosen, but its use must be correctly set up. \citet{Spada2018} first proposed to correct the prescribed entropy with an offset that corresponds to a different choice of entropy integration constant between different EoS tables, while \citet{Spada2021} proposed to only correct the prescription for the influence of a different chemical composition between the one used in the evolution and in the atmosphere model. We demonstrate in this work that both corrections must be used.

    In order to compute entropy-calibrated models, the steps are the following:
    \begin{description}
        \item[1.] First compute a standard (i.e. not with entropy-calibration) model of reference. We advocate to use a solar model as reference because it is the star for which we have the best observational constraints. This \corr{standard} model must implement exactly the same physical ingredients as the future EC model (same convection model, same opacity table, etc.).
        \item[2.] Compute the entropy offset between the \corr{standard} model and the value it should have according to a given prescription. Of course this value changes for a different mathematical formulation and a different set of coefficients.
        \item[3.] Knowing the offset value, the EC model can be computed. Each time the entropy prescription is evaluated, the resulting value must be corrected with the offset and for the different chemical composition.
    \end{description}
    This work also describes the traps that should be avoided when implementing the EC method in a stellar evolution code.

    Our implementation in \cesampp is tested with the solar model. Then we model the \acen binary system and compare the results with those from the original implementation of the method in the YREC code \citep{Spada2018,Spada2019,Spada2021}. We used a physical description as close as possible to the one used in the YREC models and performed an optimization based on the Levenberg-Marquardt algorithm. The use of the entropy calibration ties $\alpha$ to the physics of near-surface convection as modelled by RHD simulations. This consequence has many advantages. First, having a value of $\alpha$ that changes with evolution is expected because there is no reason to think that the convection should keep the same properties as a star experiences strong structural modifications. Then, since with the entropy calibration the modeller does not have to provide a value for $\alpha$, one can remove it from the set of adjustable parameters in an optimization process. This can have two effects. Either it just reduces the set of parameters and therefore facilitates the optimization. Or instead of $\alpha$, another free parameter can be adjusted in order to constrain another physical process taking place in a stellar model. EC modeling not only leads to different results than standard modeling, but also improves the physical description of convective envelope. However, it must be stressed that we do not expect any improvement of the seismic quality (e.g. a reduction of the surface effect) of stellar models by using EC models. Indeed, reliable stellar oscillations strongly depend on the modelling of the atmosphere which relies on the right choices for the opacities, the $\ttau$ relation and, the convective formalism. EC models only bring stronger constraints on the free parameters of the latter.

    Despite the numerical differences of the two codes, and a different choices of entropy prescriptions and corrections, \cesampp was able to find optimal models of \acen very similar to those obtained with YREC. About this comparison, two important observations can be made: (1) the discrepancies between standard and EC models are larger for \acen B, which is the less massive star in the system; (2) the depths of the convective envelopes are slightly different between standard and EC models. The first observation may be explained by the fact that the Hayashi track of \acen B was computed outside the range of definition of the M13. In addition, a preliminary study (presented in a forthcoming paper) suggests that the entropy of the adiabat of M dwarfs stars are not fitted accurately by the prescriptions presented in this paper. This point emphasizes the fact that entropy prescriptions should not be used outside there range of definition. A study focused on PMS and M Dwarf stars will be the topic of a future work. The second observation may be of strong importance for the understanding of chemical and angular momentum transport because the location of the convective envelope boundary has a strong impact on it. This also should be the focus of our attention in future studies.

    \begin{acknowledgements}
    We are very grateful to Federico Spada for all the help he provided and for stimulating discussions. L.M. and L.G. acknowledge support from the Max Planck Society (MPG) under project “Preparations for PLATO Science” and from the German Aerospace Center (DLR) under project "PLATO Data Center".  L.M. acknowledges support from Agence Nationale de la Recherche (ANR) grant ANR-21-CE31-0018. A.S. acknowledges grants from the Spanish program Unidad de Excelencia Mar\'{i}a de Maeztu CEX2020-001058-M, 2021-SGR-1526 (Generalitat de Catalunya), and support from ChETEC-INFRA (EU project no. 101008324). J.K. acknowledges support from European Social Fund (project No 09.3.3-LMT-K-712-19-0172) under grant agreement with the Research Council of Lithuania. Our study was partly supported by the European Union’s Horizon 2020 research and innovation program under grant agreement no. 101008324 (ChETEC-INFRA). H.-G.L. acknowledges financial support by the Deutsche Forschungsgemeinschaft (DFG, German Research Foundation) -- Project-ID 138713538 -- SFB 881 ("The Milky Way System", subproject A04).
    \end{acknowledgements}

    \nocite{OSM}
    \bibliographystyle{aa}
    \bibliography{biblio}

    \begin{appendix}

        \onecolumn

        \section{CIFIST grid for $\feh = 0.0$}
        \label{app:grid}

        \begin{table*}[!ht]
            \caption{Main characteristics of atmosphere models with solar metallicity in the CIFIST grid (the complete table is available at CDS \url{http://cds.u-strasbg.fr}. The entropy $\srhd$ is the input entropy of the model, i.e. the entropy of the inflow. Quantities with the subscript 'bot' are averages at the bottom of the simulation domain.}
            \label{table:cifist0}
            \centering
            \begin{tabular}{ l l l l l l l }
                \hline\hline
                $\teff$       &  $\logg$    & $\srhd$      & $s_{\rm bot}$ & $T_{\rm bot}$ & $\log_{10} p_{\rm bot}$ & $\rho_{\rm bot}$ \\
                $[\mbox{K}]$  &  $[\gunit]$ & $[\sunit]$  & $[\sunit]$    & [K]           & $[\punit]$              & $[10^{-5}\rhounit]$     \\
                \hline
                $6234.05$     & $      4$   & $ 2.167$    & $2.16394$     & $25382.0$     & $7.20509$               & $0.51801$       \\
                $4017.85$     & $    1.5$   & $ 2.347$    & $2.33927$     & $17506.8$     & $5.92127$               & $0.04109$      \\
                $6485.56$     & $      4$   & $ 2.377$    & $2.37243$     & $30518.6$     & $6.97112$               & $0.23571$       \\
                $6724.63$     & $   4.25$   & $ 2.366$    & $2.36573$     & $100759$      & $8.52705$               & $2.44203$        \\
                $5926.77$     & $      4$   & $ 1.992$    & $1.99046$     & $24617.6$     & $7.53879$               & $1.23210$        \\
                $5488.30$     & $    4.5$   & $ 1.693$    & $1.69257$     & $24230.8$     & $8.12577$               & $5.66237$        \\
                $4968.68$     & $    2.5$   & $2.3656$    & $2.33936$     & $15693.8$     & $5.57400$               & $0.02156$      \\
                $4924.10$     & $    3.5$   & $ 1.827$    & $1.82572$     & $20075.0$     & $7.33025$               & $1.09334$        \\
                $6233.07$     & $    4.5$   & $ 1.893$    & $1.89205$     & $26068.3$     & $7.88415$               & $2.64366$        \\
                $4774.62$     & $    3.2$   & $ 1.878$    & $1.86767$     & $14170.1$     & $6.05283$               & $0.09360$      \\
                $5884.96$     & $    3.5$   & $ 2.289$    & $2.28402$     & $24098.6$     & $6.79680$               & $0.20937$       \\
                $6456.55$     & $    4.5$   & $ 1.992$    & $1.99047$     & $26268.5$     & $7.68594$               & $1.58967$        \\
                $6431.15$     & $   4.25$   & $ 2.143$    & $2.14145$     & $29361.6$     & $7.54681$               & $0.96039$       \\
                $5226.43$     & $   4.25$   & $ 1.704$    & $1.70357$     & $23381.4$     & $8.00951$               & $4.52138$        \\
                $4954.90$     & $      4$   & $ 1.703$    & $1.70253$     & $22374.0$     & $7.89403$               & $3.68594$        \\
                $4980.91$     & $    4.5$   & $ 1.613$    & $1.61224$     & $19202.3$     & $7.66689$               & $2.85028$        \\
                $4476.99$     & $      4$   & $ 1.623$    & $1.62227$     & $18236.3$     & $7.48812$               & $2.01844$        \\
                $5537.34$     & $4.43933$   & $ 1.718$    & $1.71374$     & $15845.2$     & $6.81084$               & $0.49295$       \\
                $6084.98$     & $4.43933$   & $ 1.862$    & $1.85267$     & $15847.9$     & $6.50204$               & $0.22516$       \\
                $5475.94$     & $      4$   & $ 1.827$    & $1.82586$     & $22047.9$     & $7.59745$               & $1.77328$        \\
                $5775.01$     & $4.43933$   & $1.7734$    & $1.76747$     & $15828.4$     & $6.68128$               & $0.35625$       \\
                $5774.02$     & $4.43933$   & $1.7807$    & $1.77406$     & $15702.4$     & $6.63837$               & $0.32537$       \\
                $6227.31$     & $      4$   & $ 2.167$    & $2.16403$     & $25389.2$     & $7.20546$               & $0.51826$       \\
                $4476.44$     & $    2.5$   & $ 2.042$    & $2.03857$     & $18263.8$     & $6.63023$               & $0.22395$       \\
                $3963.69$     & $    4.5$   & $ 1.473$    & $1.47209$     & $13568.4$     & $7.10341$               & $1.32621$        \\
                $5866.50$     & $    4.5$   & $ 1.777$    & $1.77645$     & $25857.8$     & $8.10968$               & $4.76272$        \\
                $4510.49$     & $    4.5$   & $ 1.553$    & $1.55222$     & $16874.8$     & $7.44482$               & $2.11165$        \\
                $5432.51$     & $    3.5$   & $ 2.022$    & $2.01999$     & $22513.4$     & $7.25906$               & $0.71921$       \\
                \hline
            \end{tabular}
        \end{table*}

        \twocolumn

        \section{New coefficients for entropy functional representations}
        \label{app:coeff}

        \begin{table}[h!]
            \caption{Coefficients for fit to CIFIST entropies at $\feh = 0.0$ to the functional form, Eq. \eqref{eq:s_lud} by \citet{Ludwig1999}.}
            \label{table:coeff_l99}
            \centering
            \begin{tabular}{l l }
                \hline\hline
                $a_0$      &   $1.67199769$    \\
                $a_1$      &   $0.10100425$    \\
                $a_2$      &   $1.53800252$    \\
                $a_3$      &   $-1.41784705$   \\
                $a_4$      &   $0.10501243$    \\
                $a_5$      &   $-0.14985532$   \\
                \hline
            \end{tabular}
        \end{table}

        \begin{table}[h!]
            \caption{Coefficients for fit to CIFIST entropies to the functional form, Eq. \eqref{eq:s_mag} by \citet{Magic2013}.}
            \label{table:coeff_m13}
            \centering
            \begin{tabular}{l l l l }
                \hline\hline
                           & $i = 0$         & $i = 1$          & $i = 2$          \\
                \hline
                $a_i$        & $ 1.679013$ & $ 0.051669$  & $ 0.0080435$ \\
                $b_i$        & $ 0.106697$ & $-0.014053$  & $ 0.0049438$ \\
                $c_i$        & $-0.155069$ & $-0.018509$  & $-0.0028387$ \\
                $d_i$        & $ 0.094921$ & $ 0.039575$  & $ 0.0065080$ \\
                $e_i$        & $ 1.592803$ & $-0.195275$  & $-0.0015400$ \\
                $f_i$        & $-1.460810$ & $ 0.094490$  & $ 0.0024214$ \\
                \hline
            \end{tabular}
        \end{table}

        \setlength{\tabcolsep}{4.75pt}
        \begin{table}[h!]
            \caption{Coefficients for fit to CIFIST entropies to the functional form, Eq. \eqref{eq:s_tan} by \citet{Tanner2016}.}
            \label{table:coeff_t16}
            \centering
            \begin{tabular}{l l l l l }
                \hline\hline
                $\feh$      & \multicolumn{1}{c}{$-3.0$} & \multicolumn{1}{c}{$-2.0$} & \multicolumn{1}{c}{$-1.0$} & \multicolumn{1}{c}{$0.0$}          \\
                \hline
                $A$         & $0.989612$                 & $0.994939$                 & $1.004997$                 & $1.000814$   \\
                $B$         &$-0.060934$                 &$-0.066989$                 &$-0.077271$                 &$-0.081859$   \\
                $s_0$       & $1.346857$                 & $1.269913$                 & $1.398723$                 & $1.457433$   \\
                $x_0$       & $3.560125$                 & $3.550494$                 & $3.518347$                 & $3.516223$   \\
                $\beta$     & $1.196213$                 & $1.027772$                 & $0.731657$                 & $1.233787$   \\
                $\tau_0$      & $0.069621$                 & $0.103149$                 & $0.092566$                 & $0.086339$   \\
                \hline
            \end{tabular}
        \end{table}

        \begin{table}[h!]
            \caption{Values of mean molecular weight for fully ionized gas in the CIFIST grid, used for the computation of the $f_\mu$ factor of Eq. \eqref{eq:fmu}.}
            \label{table:mu_val}
            \centering
            \begin{tabular}{ l l }
                \hline\hline
                $\feh$   & $\mu_{\rm RHD}$  \\
                $-4.0$   & $0.593737$       \\
                $-3.5$   & $0.593740$       \\
                $-3.0$   & $0.593748$       \\
                $-2.5$   & $0.593773$       \\
                $-2.0$   & $0.593851$       \\
                $-1.5$   & $0.594099$       \\
                $-1.0$   & $0.594884$       \\
                $-0.5$   & $0.596241$       \\
                $ 0.0$   & $0.599408$       \\
                $ 0.5$   & $0.611516$       \\
                \hline
            \end{tabular}
        \end{table}

    \end{appendix}

\end{document}